\documentclass[reprint,superscriptaddress,aps] {revtex4-2}

\usepackage{graphicx}
\usepackage{hyperref}
\hypersetup{hidelinks}
\usepackage{color}
\usepackage{autobreak}
\usepackage{dcolumn}
\usepackage{verbatim}

\usepackage{xcolor}
\usepackage{xpatch}
 \usepackage{amsmath}
\makeatletter

\begin{document}
\preprint{APS/123-QED}

\title{Heat conduction theory including phonon coherence}

\author{Zhongwei Zhang}
\email{zhongwei@iis.u-tokyo.ac.jp}
\affiliation{Institute of Industrial Science, The University of Tokyo, Tokyo 153-8505, Japan} 

\author{Yangyu Guo}
\affiliation{Institute of Industrial Science, The University of Tokyo, Tokyo 153-8505, Japan}

\author{Marc Bescond}
\affiliation{Laboratory for Integrated Micro and Mechatronic Systems, CNRS-IIS UMI 2820, The University of Tokyo, Tokyo 153-8505, Japan}

\author{Jie Chen}
\email{jie@tongji.edu.cn}
\affiliation{Center for Phononics and Thermal Energy Science, School of Physics Science and Engineering and China-EU Joint Lab for Nanophononics, Tongji University, Shanghai 200092, People's Republic of China}
             
\author{Masahiro Nomura}
\affiliation{Institute of Industrial Science, The University of Tokyo, Tokyo 153-8505, Japan}

\author{Sebastian Volz}
\email{volz@iis.u-tokyo.ac.jp}
\affiliation{Institute of Industrial Science, The University of Tokyo, Tokyo 153-8505, Japan}
\affiliation{Laboratory for Integrated Micro and Mechatronic Systems, CNRS-IIS UMI 2820, The University of Tokyo, Tokyo 153-8505, Japan}
\affiliation{Center for Phononics and Thermal Energy Science, School of Physics Science and Engineering and China-EU Joint Lab for Nanophononics, Tongji University, Shanghai 200092, People's Republic of China}

\date{\today}

\begin{abstract}

Understanding and quantifying the fundamental physical property of coherence of thermal excitations is a long-standing and general problem in physics. The conventional theory, i.e. the phonon gas model, fails to describe coherence and its impact on thermal transport. In this letter, we propose a general heat conduction formalism supported by theoretical arguments and direct atomic simulations, which takes into account both the conventional phonon gas model and the wave nature of thermal phonons. By naturally introducing wavepackets in the heat flux from fundamental concepts, we derive an original thermal conductivity expression including coherence times and lifetimes. Our theory and simulations reveal two distinct types of coherence, i.e., intrinsic and mutual, appearing in two different temperature ranges. This contribution establishes a fundamental frame for understanding and quantifying the coherence of thermal phonons, which should have a general impact on the estimation of the thermal properties of solids.

\end{abstract}

\pacs{Valid PACS appear here}
\maketitle


Phonons, i.e., quanta of vibrational waves, are commonly considered as one of the fundamental quasi-particles, simultaneously exhibiting wavelike and particlelike characteristics in nanostructured crystals or bulk materials. The wavelike behavior of phonons impacts thermal properties via coherence mechanisms, as highlighted by several pioneering \cite{ZH606,ZH608} and recent works \cite{RN1616,RN1617}. The particlelike behavior has been treated by Boltzmann transport equation (BTE) and the phonon-gas model in most solids \cite{ZH116,ziman2001,Guyer1966,McGaughey2004a,Cepellotti2016,Minnich2015}. Experiments \cite{ZH606,ZH608,Zen2014,ZH975,Hu2020a} have revealed, however, that the wave nature of thermal phonons plays a substantial role in thermal transport, as for example, in the observations of coherent thermal transport in nano-phononic crystals \cite{ZH606,ZH608,Zen2014,ZH975}. Later, theoretical and simulation studies \cite{Tian2014,Wang2014d,Mendoza2016,Hu2018,Juntunen2019} were devoted to the understanding of phonon coherence, such as the one producing band folding \cite{Maris1999,Bies2000,Maldovan2015}, but missing the particle behavior. Recently, the theoretical study \cite{Zhang2021} revealed that the realistic phonon dynamics can only be manifested if both intrinsic coherence relevant to the extension of phonon wavepackets and the particlelike behavior of thermal phonons are taken into account.

The conventional BTE also fails in complex crystals, as a pure particle picture cannot yield a complete description of thermal conductivity, such as in Tl$_{3}$VSe$_{4}$ \cite{Luo2020}. Recently, Simoncelli \textit{et al.} \cite{RN1616} developed a theory for thermal transport in glasses and complex crystals, in which the coherence between densely packed phonon branches contributes to thermal transport. A similar approach has been developed by Isaeva \textit{et al.} \cite{RN1617} as well at the same time. This mutual coherence among branches is identified as an additional phonon wave-relevant term \cite{Xia2020,Xia2020,Jain2020,Hanus2021}. The picture of this mutual coherence, that might be compared to a hopping process, however remains physically unclear. Finally, quantifying the full coherence of thermal phonons and its effect on heat conduction remains a critical issue in transport physics.

\begin{figure}[t]
\includegraphics[width=1.0\linewidth]{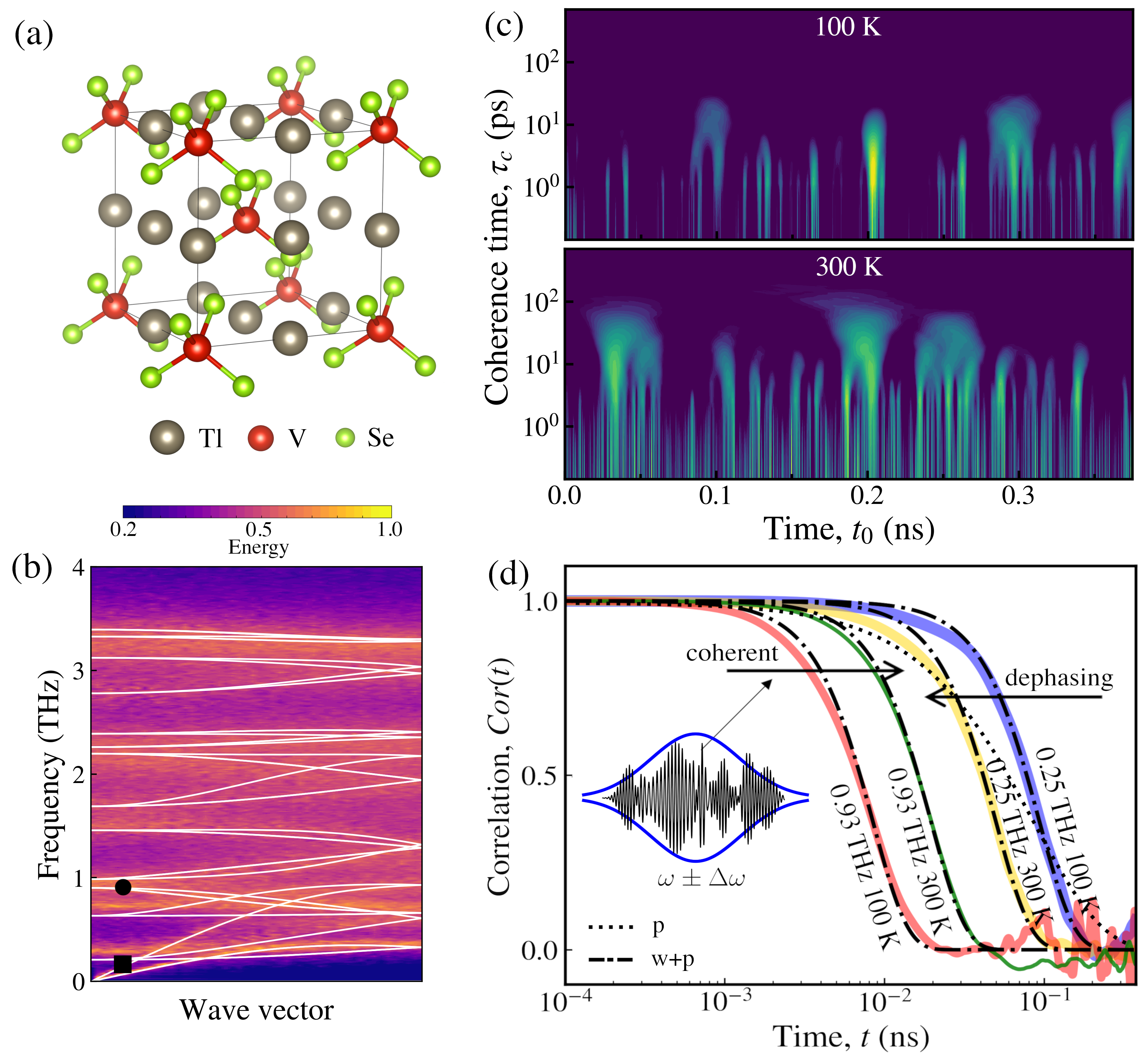}
\caption{(a) The conventional lattice of Tl$_{3}$VSe$_{4}$. (b) The phonon dispersion of Tl$_{3}$VSe$_{4}$ from lattice dynamic calculations (white lines) and room-temperature spectral energy density calculations (contour) for the conventional cell. The symbols in (b) indicate the instances of modes being analyzed, 0.25 THz (square) and 0.93 THz (circle). (c) Evolution time and coherence time dependent phonon number of Tl$_{3}$VSe$_{4}$ for the 0.93 THz mode at 100 K and 300 K. (d) Phonon decay (correlation) versus correlation time in Tl$_{3}$VSe$_{4}$ for the 0.25 and 0.93 THz modes at 100 K and 300 K. The dash-dotted lines (p+w) show the fitting by Eq. (\ref{eq13}). The dotted line (p) shows the fitting by the conventional exponential decay. The inset of (d) shows the realistic wavepacket of the mode 0.93 THz resulting from the combination of shorter wavepackets at 300 K.
}
\label{figure1}
\end{figure}

In this letter, a general heat conduction theory is proposed to establish an original expression for the thermal conductivity that includes the full coherent nature of phonon excitations. This expression involves both phonon lifetimes and coherence times. Those are obtained by tracking the real phonon dynamics and using a wavelet transform of the atomic trajectories during an equilibrium molecular dynamic (EMD) simulation. We show that the predictions of our theory yield significant differences from those of the conventional phonon-gas model, as demonstrated in the Tl$_{3}$VSe$_{4}$ case (See atomic structure in Fig.\,\ref{figure1}(a)). We find that there are two types of coherence, i.e., intrinsic and mutual, which take a critical role over different temperature regions. These conclusions open unexpected insights on the reality of thermally activated phonon modes and the importance of the diverse coherence mechanisms when assessing thermal properties.

The thermal conductivity ($\kappa$) can be calculated based on the Green-Kubo approach with the autocorrelation of the heat flux $\mathbf{S}\left ( t \right )$ as \cite{Kubo1966}

\begin{eqnarray}
\kappa =\frac{V}{3k_{B}T^{2}}\int \left \langle \mathbf{S}\left ( t \right ) \cdot  \mathbf{S}\left ( 0\right )\right \rangle dt,
\label{eqs1}
\end{eqnarray}

\noindent where $V$ corresponds to the system volume, $k_{B}$ is the Boltzmann constant and $T$ the temperature. We now define the heat flux component $\mathbf{S}_{\lambda}$ corresponding to the contribution of mode $\lambda $ to the total heat flux, i.e., $ \mathbf{S}\left ( t \right )=\sum_{\lambda} \mathbf{S}_{\lambda }\left ( t \right )$. $\lambda $ refers to a specific mode of wavevector $\mathbf{k}$ and of the branch $s$, which implies that the correlation can operate between modes of different wavevectors and of different branches. Then, the thermal conductivity can be decomposed as follows

\begin{eqnarray}
\kappa _{d}= \frac{V}{3k_{B}T^{2}}\sum_{\lambda} \int \left \langle  \mathbf{S}_{\lambda }\left ( t \right )\cdot \mathbf{S}_{\lambda}\left ( 0\right )\right \rangle dt.
\label{eqs2}
\end{eqnarray}

As discussed before by Gill-Comeau and Lewis \cite{RN1611,RN1612}, the neglected off-diagonal part corresponds to the collective excitations of phonons ($ \sim \left \langle  \mathbf{S}_{\lambda }\left ( t \right )\cdot  \mathbf{S}_{\lambda^{'}\neq \lambda}\left ( 0\right )\right \rangle$). In most solids, $\kappa_{d}$ defined in Eq. (\ref{eqs2}) is predominant and the off-diagonal terms only become significant in the hydrodynamic regime \cite{Cepellotti2016,Guo2015,Ding2018b,Zhang2020}, for instance in graphene and graphite. Therefore, the diagonal thermal conductivity only will be addressed in the following. Based on the definition of the harmonic heat flux operator \cite{ZH226}, $\mathbf{S}=\frac{1}{V}\sum_{\lambda}N_{\lambda}\hbar\omega _{\lambda}\boldsymbol{\upsilon }_{\lambda }$, the thermal conductivity can be derived as

\begin{flalign}
\kappa =  \frac{1}{3k_{B}VT^{2}}\sum_{\alpha}\sum_{\lambda} \hbar^{2}\omega _{\lambda}^{2 }\upsilon _{\lambda,\alpha }^{2} N_{\lambda }^{2}\int Cor_{\lambda}\left ( t\right )dt,
\label{eqs4}
\end{flalign}

\noindent here, $\hbar$ denotes the reduced Planck constant, $\omega _{\lambda}$ is the eigenfrequency, $\boldsymbol{\upsilon } _{\lambda}$ the group velocity of mode $\lambda$ and $\alpha$ the Cartesian coordinate. $N_{\lambda }\left ( t \right )$ is the time-dependent phonon number of mode ${\lambda }$ and $Cor_{\lambda}\left ( t\right )=\frac{\left \langle  N_{\lambda }\left ( t \right )N_{\lambda}\left ( 0\right )\right \rangle }{\left \langle  N_{\lambda }\left ( 0 \right )N_{\lambda}\left ( 0\right )\right \rangle }$ the phonon decay. Considering the classical definition of the phonon number $N_{\lambda }=\frac{k_{B}T}{\hbar\omega _{\lambda}}$, Eq. (\ref{eqs4}) reduces to $\kappa =  \frac{1}{3}\sum_{\alpha }\sum_{\lambda} C _{\mathrm{v},\lambda}^{clas} \upsilon _{\lambda,\alpha }^{2} \int Cor_{\lambda}\left ( t\right )dt$, where $C _{\mathrm{v},\lambda}^{clas}= \frac{k_{B}}{V}$ is the classical specific heat per mode. When the particlelike behavior predominates, $Cor_{\lambda}\left ( t\right )$ follows an exponential decay law with the lifetime ${\tau _{\lambda}^{p}}$ according to the conventional single-mode relaxation time (SMRT) approximation \cite{McGaughey2004a,Minnich2015}, that is $Cor_{\lambda}\left ( t\right )=e^{-\frac{t}{\tau_{\lambda}^{p} }}$. After integration, the classical particlelike thermal conductivity can be expressed as follows
 
\begin{eqnarray}
\kappa_{p} =  \frac{1}{3}\sum_{\alpha }\sum_{\lambda} C _{\mathrm{v},\lambda}^{clas} \upsilon _{\lambda,\alpha}^{2} \tau_{\lambda}^{p}.
\label{eqs6}
\end{eqnarray}

Equation (\ref{eqs6}) is analogous to the commonly used Peierls-Boltzmann formula \cite{ziman2001}. However, as the coherence effect increases, a correction should be considered in the description of the phonon decay by including the modal coherence time (${\tau_{\lambda}}^{c}$) \cite{Zhang2021} as follows  

\begin{eqnarray}
Cor_{\lambda}\left ( t\right )= e^{-\frac{t}{2\tau_{\lambda}^{p} }}e^{-4ln2\frac{t^{2}}{{\tau_{\lambda}}^{c2}}}.
\label{eq13}
\end{eqnarray}
 
\noindent The second gaussian term originates from the interference between different modes, expressing coherence effects in the phonon dynamics. Consequently, by integrating Eq. (\ref{eq13}), the complete thermal conductivity ($\kappa _{p+w}$) including the wavelike behavior can be expressed as 

\begin{eqnarray}
\kappa _{p+w} =  \frac{1}{3}\sum_{\alpha }\sum_{\lambda} C _{\mathrm{v},\lambda}^{clas} \upsilon _{\lambda,\alpha}^{2}\sqrt{\frac{\pi }{4ln2}}\tau _{\lambda}^{c} e^{\frac{{\tau _{\lambda}^{c2}} }{128ln2{\tau_{\lambda}^{p2}} }}.
\label{eqs15}
\end{eqnarray}

When setting $\tau_{\lambda}^{p}=\tau_{\lambda}^{c}$, Eq. (\ref{eqs15}) reduces to the conventional Peierls-Boltzmann formula with a corrective coefficient $\approx$ 1.07. This consideration reveals that equality between coherence time and lifetime \cite{Latour2017} is underlying the traditional phonon gas model.

In realistic systems, the scatterings are diverse and can also be incorporated into the model of Matthiessen

\begin{eqnarray}
\frac{1}{\tau _{\lambda }^{total}}=\frac{1}{\tau _{\lambda }^{ph-ph}}+\frac{1}{\tau _{\lambda }^{isotope}}+\frac{1}{\tau _{\lambda }^{boundary}}.
\label{eqs16}
\end{eqnarray}

\noindent where, $\tau _{\lambda }^{ph-ph}$ refers to the phonon decay time for $\tau_{\lambda}^{p}$ and $\sqrt{\frac{\pi }{4ln2}}\tau _{\lambda}^{p} e^{\frac{{\tau _{\lambda}^{c2}} }{128ln2{\tau_{\lambda}^{p2}} }}$ in Eq. (\ref{eqs6}) and Eq. (\ref{eqs15}), respectively. The calculations of the isotope scattering time $\tau _{\lambda }^{isotope}$ \cite{Tamura1983} and of the boundary scattering time $\tau _{\lambda }^{boundary}$ \cite{Casimir1938} follow the reference \cite{Li2014}. Only the isotope scattering is intrinsically considered in our following calculations. Quantifying the contribution of coherence as proposed in Eq. (\ref{eqs15}) requires the knowledge of lifetimes and coherence times. These quantities are accessible by a wavelet approach \cite{Zhang2021,SM}.
 
 The developed model is then applied to the complex crystal Tl$_{3}$VSe$_{4}$ (See atomic structure in Fig.\,\ref{figure1}(a)). In order to collect accurate phonon information from MD simulations, a machine learning potential (MLP) \cite{Shapeev2016,Novikov2021} is adopted after being trained by $ab$ $initio$ MD simulations \cite{Kresse1993,Kresse1996,KRESSE199615,Kresse1994}. In the Supplementary Materials (SM) \cite{SM}, the training process and the verification of the MLP are described \cite{Li2014,Li2012,phonopy,Ouyang2020}. All MD simulations using the MLP are performed with the LAMMPS package \cite{Plimpton1995}, which is mainly used for calculating MD thermal conductivity \cite{Kubo1966,Torii2008} for comparison (See Sec.s2.3 in \cite{SM}), and for extracting the atomic information for subsequent wavelet transform calculations (See Sec.s3 in \cite{SM} for details).

By applying the wavelet transform approach, we can firstly obtain the evolution time and coherence time dependent phonon number $N\left ( t_{0},\tau _{\lambda}^{c}  \right )$. Increase temperature usually dephases the correlation of phonon waves or suppresses the coherent/wavelike behaviour, resulting in a decrease in coherence time. The coherence time of the 0.25 THz mode in Tl$_{3}$VSe$_{4}$, for example, decreases when the temperature rises from 100 K to 300 K (See Fig. S5). This dephasing trend has already been seen in bulk graphene \cite{Zhang2021} and superlattices \cite{ZH606,Chen2016a,Hu2018} before which corresponds to a loss of `intrinsic' coherence. However, for high frequency phonons in Tl$_{3}$VSe$_{4}$, e.g., at 0.93 THz shown in Fig.\,\ref{figure1}(c), the coherence time reversely increases with temperature, as evidenced by the expanded `clouds' and their shift into a higher coherence time regime. These `clouds' can be understood as the coherence effect between the modes of different branches in the phonon dispersion (Fig.\,\ref{figure1}(b)), as demonstrated before by Simoncelli \textit{et al.} \cite{RN1616} and Isaeva \textit{et al.} \cite{RN1617}. The clouds with long coherence times are connected to the packets with small coherence times, as if the former were generated by the latter.

The abundance of small wavepackets with shorter lifetimes is partially originating from the temperature induced phonon-phonon scattering. Since lifetime is reversely proportional to the frequency broadening (i.e. $\Delta \omega$), the enlarged linewidth of the dispersion curves associated with reduced frequency intervals between branches promote smaller wavepackets as illustrated by Fig.\,\ref{figure1}(b-c). In a subsequent stage, due to their large number, those small wavepackets have higher probability to establish a phase relation between themselves, which results in the generation of wavepackets with large coherence times, referred to as `clouds' above (See the inset of Fig.\,\ref{figure1}(d)). This argument does not only support the model of Simoncelli \textit{et al.} \cite{RN1616}, but also offers a physical picture of the mutual coherence for phonons in complex crystals. In addition, this effect should also be amplified as temperature further rises due to the increase of branches broadening from phonon-phonon scatterings. Note that these tendencies are also respectively observed in other low or high frequency modes.

Coherence is expected to have critical impact on the phonon decay and its propagation. The phonon decay can be extracted from the autocorrelation function of the phonon number $N\left ( t_{0},\tau _{\lambda}^{c}  \right )$, as displayed in Fig.\,\ref{figure1}(d). The details of the computations are provided in \cite{SM}. When including the coherence effect, the phonon decay should follow the generalized law of Eq. (\ref{eq13}) as we demonstrated before \cite{Zhang2021}. From the fitted particle-related decay (dotted line) to the coherence corrected decay (dash-dotted line) of Fig.\,\ref{figure1}(d), the coherence appears as a delay to the conventional exponential decay. In addition, the phonon decay of the low frequency 0.25 THz mode shows the temperature effects on the coherence dephasing and reduction of phonon lifetimes. In contrast, for the high frequency 0.93 THz at which branch density is high, the increase of temperature contributes to coherence, producing a further delay in phonon decay. The inset of Fig.\,\ref{figure1}(d) shows how small wavepackets combine into a large one in this frequency range. By fitting this phonon decay with the generalized decay law of Eq. (\ref{eq13}), we obtain $\tau _{\lambda}^{p}$ and $\tau _{\lambda}^{c}$ for different modes as reported in Fig. S6. The fitting details can be found in \cite{SM}. A $\mathbf{q}$-grid of $16\times16\times16$ is applied to analyze the modes in the full Brillouin zone, this mesh is yielding converged results in BTE based estimations. Importantly, $\tau _{\lambda}^{c}$ becomes larger than $\tau _{\lambda}^{p}$ in the region of relatively high frequencies, because of the coherence effect (See Fig. S6).

\begin{figure}[t]
\includegraphics[width=1.0\linewidth]{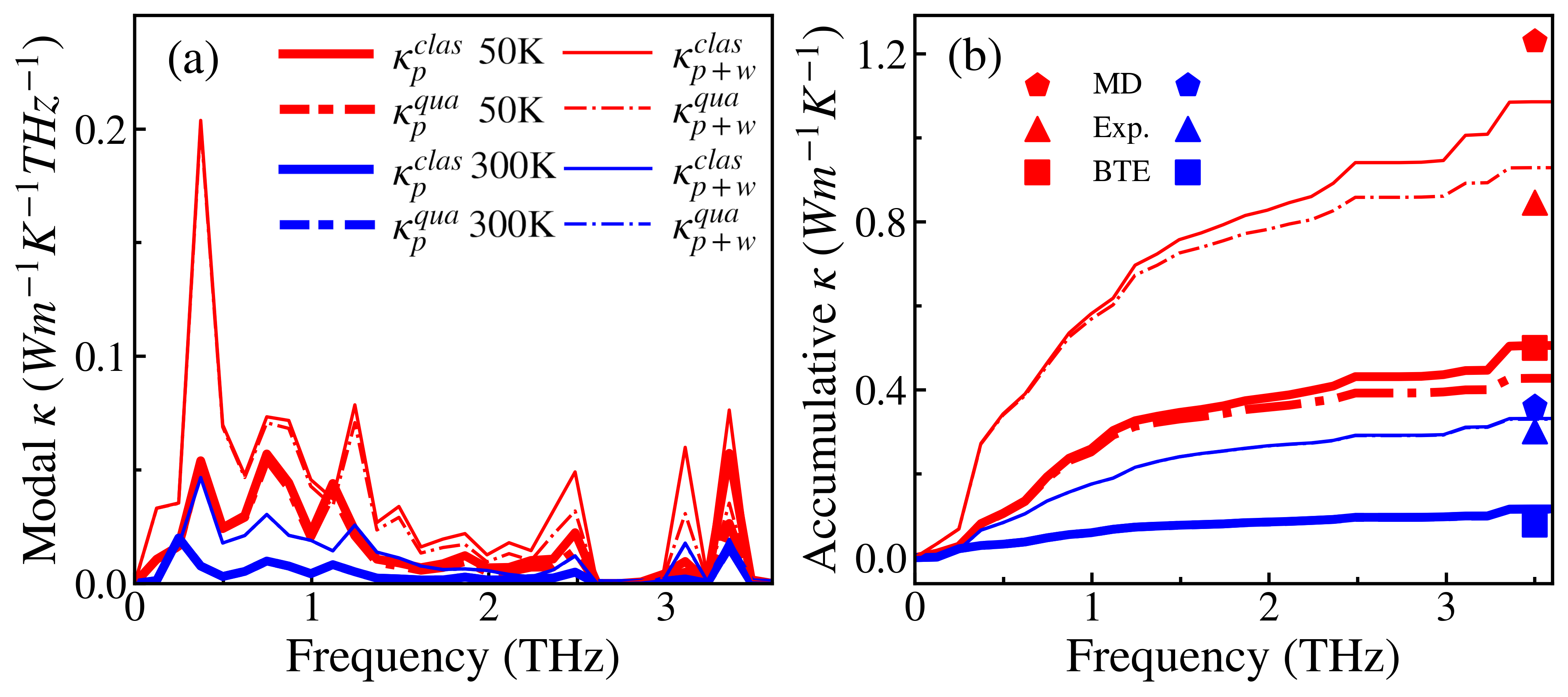}
\caption{Modal thermal conductivity of Tl$_{3}$VSe$_{4}$. (a) Modal classical and quantum thermal conductivities of Tl$_{3}$VSe$_{4}$ at 50 K and 300 K. (b) Accumulative classical and quantum thermal conductivities of Tl$_{3}$VSe$_{4}$ at 50 K and 300 K. The symbols refer to experimental results (Exp.) \cite{Mukhopadhyay2018}, Boltzmann transport equation results (BTE) \cite{Jain2020} and our molecular dynamic simulations (MD). 
}
\label{figure2}
\end{figure}

 Due to the predominance of phonons of low frequencies, the effect of quantum population on the transport properties of Tl$_{3}$VSe$_{4}$ is limited. The thermal conductivities of Eqs. (\ref{eqs6}) and (\ref{eqs15}) are expressed with the classical population (i.e. $\kappa_{p/p+w}^{clas}$), which can be quantum corrected (i.e. $\kappa_{p/p+w}^{qua}$) by replacing $C _{\mathrm{v},\lambda}^{clas}$ with the quantum specific heat $C _{\mathrm{v},\lambda}^{qua}$, here $C _{\mathrm{v},\lambda}^{qua}=\frac{k_{B}exp\left ( x  \right )}{V}\left [\frac{x}{exp\left ( x  \right )-1} \right ]^{2} $ with $x=  \frac{\hbar \omega _{\lambda} }{k_{B}T} $. At low temperatures, Fig. \,\ref{figure2}(a) reports that the delay effect on phonon decay due to the intrinsic coherence leads to substantial contributions to thermal conductivity in the full frequency range. The effect of intrinsic coherence should be weakened by increasing temperature. The wavelike contributions are still prominent at 300 K and the additional components are migrating to the region after 0.25 THz where the dense branches and mutual coherence emerge. Obviously, coherence can correct thermal conductivities by incorporating the wavelike contribution, and coherence behaviors are different at low temperatures (intrinsic coherence) and at high temperatures (mutual coherence). 
 
 The accumulative $\kappa$ in Fig.\,\ref{figure2}(b) further evidences the importance of wavelike contributions to thermal conductivities in Tl$_{3}$VSe$_{4}$. Here, the BTE results are from Jain \cite{Jain2020}, in which a temperature dependent effective potential (TDEP) is considered. As all scatterings and temperature effects are included in MD simulations, our predicted $\kappa _{p}^{clas/qua}$ are quite close to the BTE results, indicating the validity of the particle description in our model. When including the wavelike behaviour of thermal phonons, the corrected $\kappa _{p+w}^{clas/qua}$ approximate to both the experimental measurements \cite{Mukhopadhyay2018} and the direct MD simulations, especially at 300 K. The limited discrepancies at 50 K are mostly originating from the classical population of phonons in MD simulation, which overestimates phonon-phonon scatterings while underestimating $\kappa _{p}^{clas}$ and $\kappa _{p+w}^{clas}$ with suppressed $\tau_{\lambda}^{p}$ and $\tau_{\lambda}^{c}$.

\begin{figure}[b]
\includegraphics[width=1.0\linewidth]{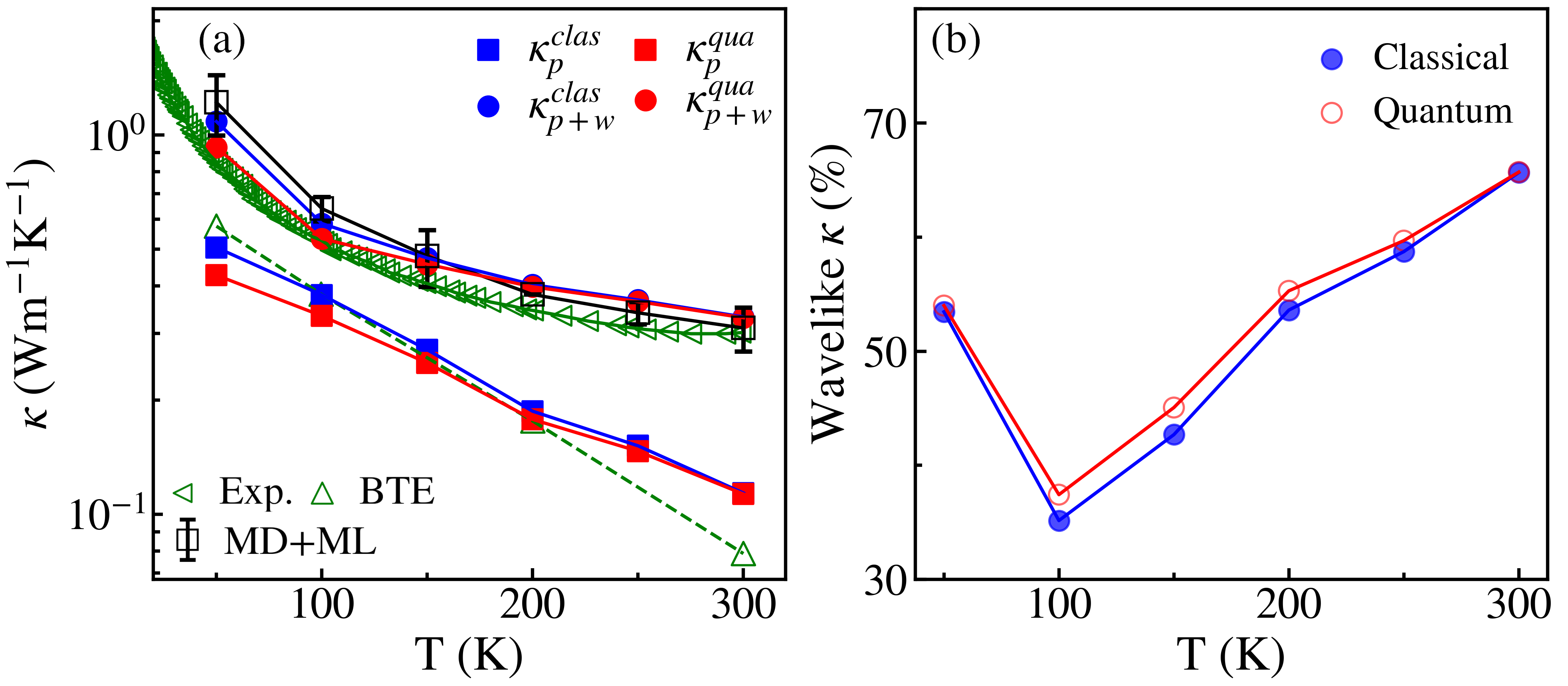}
\caption{Comparison of thermal conductivities of Tl$_{3}$VSe$_{4}$ from different approaches. (a) The comparison of thermal conductivities of Tl$_{3}$VSe$_{4}$ versus temperature. The left oriented triangles are experimental results (Exp.) \cite{Mukhopadhyay2018}, the upward triangles refer to BTE results \cite{Jain2020} and the squares correspond to our machine learning based MD simulations (MD+ML). Full symbols denote the models of Eq. (\ref{eqs6}) (squares) and of our theory of Eq. (\ref{eqs15}) (circles) in the classical (blue) and the quantum (red) approximations. (b) The additional contribution of coherence to thermal conductivities of Tl$_{3}$VSe$_{4}$ versus temperature.
}
\label{figure3}
\end{figure}

For overall comparison, thermal conductivities based on our direct MD simulations, experimentally measured values \cite{Mukhopadhyay2018} and the BTE calculations with TDEP \cite{Jain2020} are summarized in Fig.\,\ref{figure3}(a). MD simulations implicitly include all orders of anharmonicity for phonon-phonon scattering as well as both the wavelike and particlelike behaviors of phonons. Consequently, EMD simulations include the real phonon dynamics and well coincide with the experimental $\kappa$, indicating the accuracy of our MLP. The computed particlelike thermal conductivities ($\kappa _{p}^{clas/qua}$) agree well with the BTE results, but remains lower than the  experimental measurements, indicating the failure of BTE in capturing phonon coherence of Tl$_{3}$VSe$_{4}$. As we further include the wavelike contribution by applying the revised heat conduction law of Eq. (\ref{eqs15}), the thermal conductivities significantly increase. The wavelike corrected values $\kappa _{p+w}^{clas/qua}$ agree well with the experimental measurements and MD simulations in the full temperature region. This comparison indicates that the proposed model is sufficient in predicting thermal transport by including the wavelike features of phonons.
 
The degree of coherence correction to thermal conductivity is further estimated in Fig.\,\ref{figure3}(b) from $\frac{\kappa _{p+w}^{clas/qua}-\kappa _{p}^{clas/qua}}{\kappa _{p+w}^{clas/qua}}\times 100\%$. Interestingly, a non-monotonic dependence of the coherence contribution as a function of temperature is observed, which agrees well with the above demonstration of the coexistence of two types of coherence. At room-temperature, the wavelike correction reaches 66 $\%$. The trend of the correction with temperature coincides with the discrepancy between the prevailing BTE theory and the real phonon dynamics observed in the experiments and MD simulations of Fig.\,\ref{figure3}(a), revealing that the wavelike behavior becomes more significant at low and high temperatures. In Ref. \cite{Jain2020}, Jain found that when including coherence as proposed by Simoncelli \textit{et al.} \cite{RN1616}, a conductivity close to the experimental measurements could be obtained but still smaller, especially at low temperatures where mutual coherence disappears and intrinsic coherence becomes important (See the comparison in \cite{SM}). The used Allen and Feldman (AF) model \cite{Allen1989,Allen1993} and Cahill-Watson-Pohl (CWP) model \cite{Cahill1989,Cahill1992}, as well as the theory from \cite{RN1616}, are based on the weakly propagating modes and the inter-modes `hopping' additionally contributing to thermal transport in complex crystals or disordered systems. The discovered mutual coherence offers direct evidence on how modes interact and affect the thermal transport. The above comparisons between thermal conductivities demonstrate the remaining issues of prevailing theories and indicate the ability of our model to capture both wavelike and particlelike pictures of phonons in thermal transport. The agreement with measurements especially provides a satisfactory test of our theory based on experimental data.

Coming back to the fundamental definition of the heat flux, we have established a new expression of the thermal conductivity including phonon coherence via coherence time. The estimation of this generalized thermal conductivity was implemented by introducing a wavelet transformation of MD quantities, which offers the coherence time of phonon excitations as well as their lifetimes. This methodological treatment of coherence in heat conduction has hence unraveled the rich content of coherent effects on thermal transport. This previously uncharted information has led to quantitative estimation of the impact of coherence responsible for a 66 $\%$ increase of thermal conductivity at room-temperature for complex crystal Tl$_{3}$VSe$_{4}$. In addition, two distinct types of coherence are observed in our explorations, including intrinsic coherence and mutual coherence, which can be simultaneously modeled by our heat conduction theory. In short, we proposed a novel paradigm to describe the full coherence of phonon heat carriers, which has a global repercussion in the assessment of the thermal properties in nanostructures as well as in bulk materials.

\section*{\label{sec:level1}Acknowledgments}

This work is partially supported by CREST JST (No. JPMJCR19I1 and JPMJCR19Q3). This research used the computational resources of the Oakforest-PACS supercomputer system, The University of Tokyo. This project is also supported in part by the grants from the National Natural Science Foundation of China (Grant Nos. 12075168 and 11890703), and Science and Technology Commission of Shanghai Municipality (Grant No. 19ZR1478600). 
 
\bibliographystyle{apsrev4-2}
\bibliography{library}

\end{document}



\title{Supplementary Material for `Heat conduction theory including phonon coherence'}

\author{Zhongwei Zhang}
\email{zhongwei@iis.u-tokyo.ac.jp}
\affiliation{Institute of Industrial Science, The University of Tokyo, Tokyo 153-8505, Japan}

\author{Yangyu Guo}
\affiliation{Institute of Industrial Science, The University of Tokyo, Tokyo 153-8505, Japan}

\author{Marc Bescond}
\affiliation{Laboratory for Integrated Micro and Mechatronic Systems, CNRS-IIS UMI 2820, The University of Tokyo, Tokyo 153-8505, Japan}

\author{Jie Chen}
\email{jie@tongji.edu.cn}
\affiliation{Center for Phononics and Thermal Energy Science, School of Physics Science and Engineering and China-EU Joint Lab for Nanophononics, Tongji University, Shanghai 200092, People's Republic of China}
             
\author{Masahiro Nomura}
\affiliation{Institute of Industrial Science, The University of Tokyo, Tokyo 153-8505, Japan}

\author{Sebastian Volz}
\email{volz@iis.u-tokyo.ac.jp}
\affiliation{Institute of Industrial Science, The University of Tokyo, Tokyo 153-8505, Japan}
\affiliation{Laboratory for Integrated Micro and Mechatronic Systems, CNRS-IIS UMI 2820, The University of Tokyo, Tokyo 153-8505, Japan}
\affiliation{Center for Phononics and Thermal Energy Science, School of Physics Science and Engineering and China-EU Joint Lab for Nanophononics, Tongji University, Shanghai 200092, People's Republic of China}
  
\maketitle


\section{Construction of machine learning potential}

\subsection{$Ab$ $initio$ MD simulations}

The VASP code \cite{Kresse1993,Kresse1996,KRESSE199615} is used with potpaw-PBE.54 \cite{Kresse1994} pseudopotentials to perform $ab$ $initio$ molecular dynamic (AIMD) simulations to collect the energies, atomic force, and stress as train datasets for the training of the machine learning potential (MLP). The system size of Tl$_{3}$VSe$_{4}$ for AIMD calculations is 3×3×3 primitive cells containing 216 atoms. The cutoff energy of 400 eV is used. A 2×2×2 $\Gamma$-centered grid of k points in the irreducible Brillouin zone is used. To fully consider the effect of temperature on structure, AIMD is performed with isothermal-isobaric (NPT) ensemble from 0 K to 1000 K. The time step is 0.5 fs. The 2000 and 4000 AIMD atomic configurations and the corresponding atomic forces, total energy, and stress are used to train machine learning potential \cite{Shapeev2016,Novikov2021}.

\subsection{Machine learning potential}
 
\begin{figure}[h]
\includegraphics[width=1.0\linewidth]{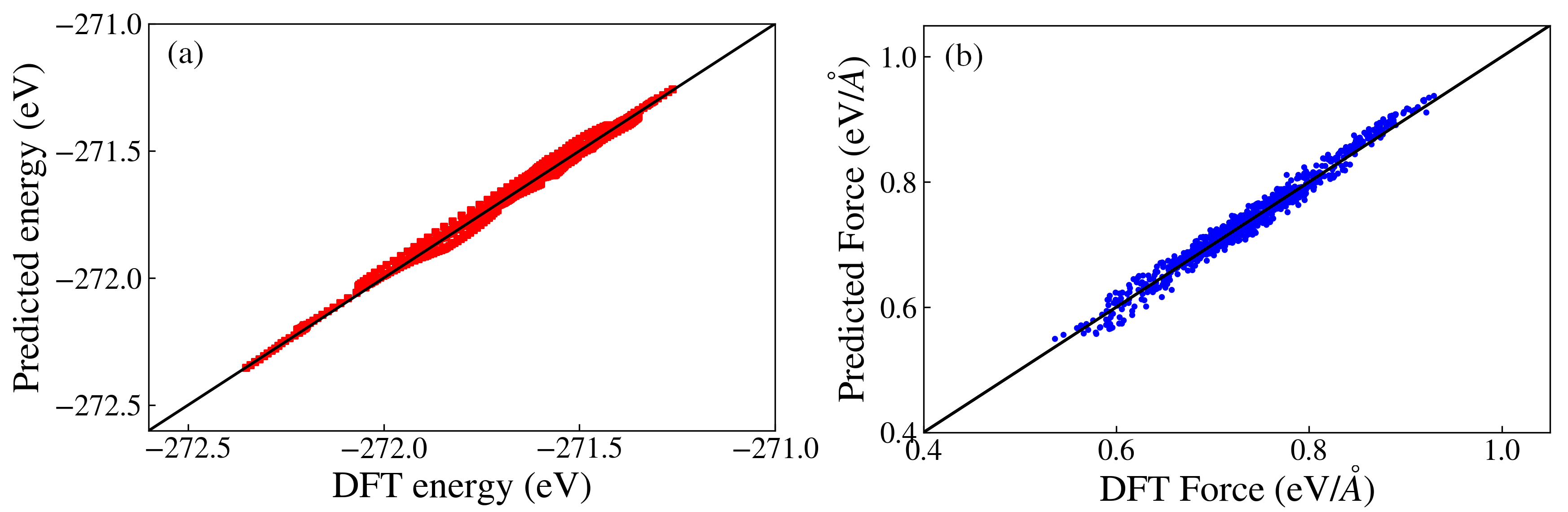}
\caption{(a) The comparison of potential energy between DFT calculation and MLP based prediction. (b) The comparison of averaged forces between DFT calculation and MLP based prediction. The forces are averaged over different atoms. 200 samples are used for the testing.
}
\label{figs1}
\end{figure}

 We employ a moment tensor potential developed by A. V. Shapeev \cite{Shapeev2016,Novikov2021} as a machine learning potential MLP model to describe the interatomic interactions in Tl$_{3}$VSe$_{4}$. The details of the training process can be found in references \cite{Shapeev2016,Novikov2021,Ouyang2020}. With the trained MLP and an interface to LAMMPS software \cite{Novikov2021,Plimpton1995}, we would be able to do further lattice dynamic and molecular dynamic calculations. To begin with, the trained MLP is tested by comparing the energies and atomic forces predicted by DFT and MLP predictions, which reveals a good agreement between the two calculations (See Fig.\,\ref{figs1}). During the calculations, we find that the lattice constant is optimized at the value of 7.892 $\AA$ which agrees well with the DFT result of 7.887 $\AA$ and also with the experimental value of 7.904 $\AA$ \cite{Mukhopadhyay2018}.

\section{Benchmarks of MLP}

\subsection{Phonon dispersion}

\begin{figure}[h]
\includegraphics[width=0.5\linewidth]{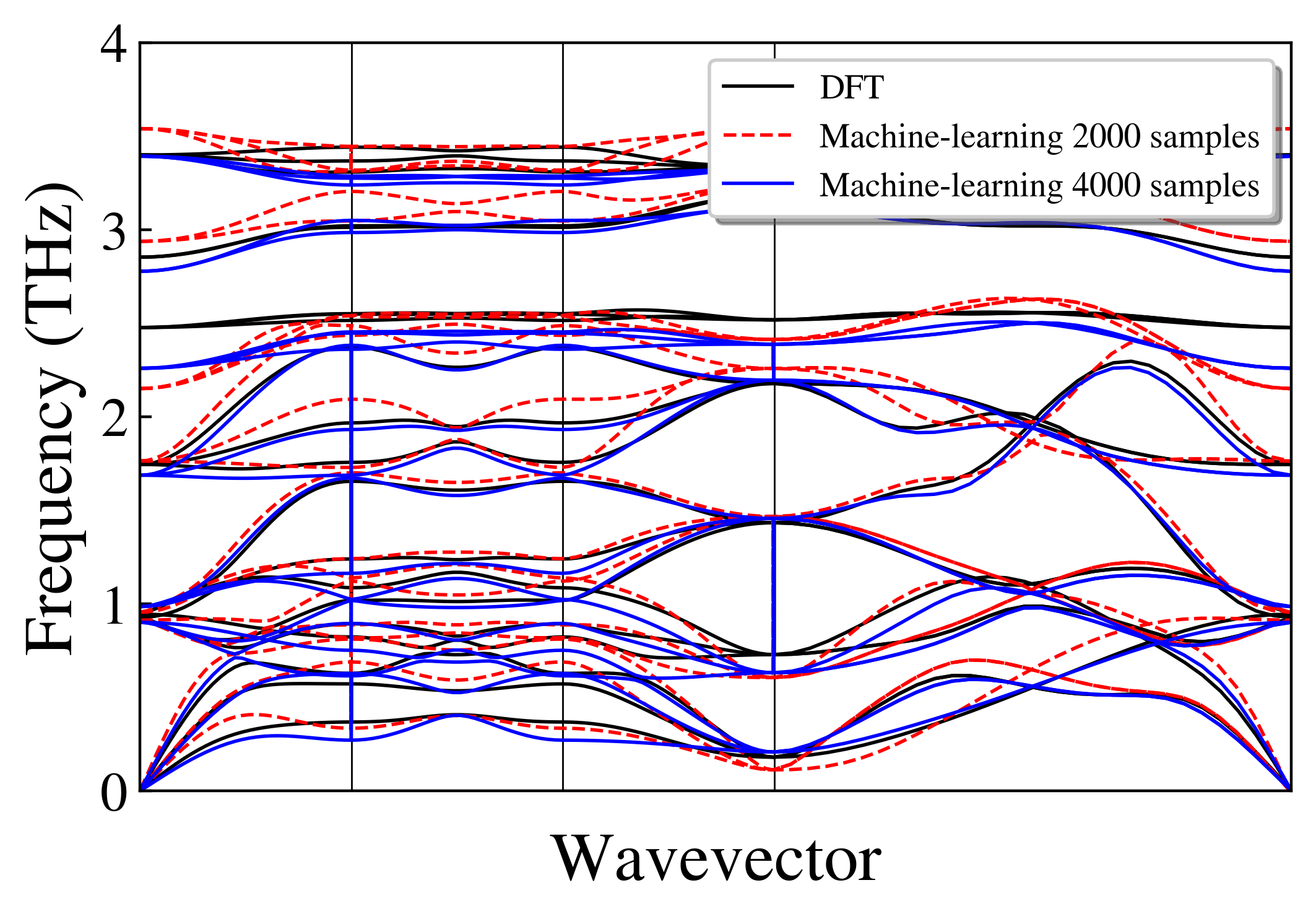}
\caption{The comparison of phonon dispersion from DFT calculation and MLP prediction with different training set.
}
\label{figs2}
\end{figure}

Then we compare the phonon dispersion from DFT and MLP calculations. The phonon dispersion is calculated by using the PHONOPY code \cite{phonopy} with 3×3×3 primitive cells. As shown in Fig.\,\ref{figs2}, the phonon dispersion from MLP is in agreement with the DFT calculations, indicating the accuracy of MLP in terms of harmonic properties prediction. In addition, by increases the training set from 2000 samples to 4000 samples, the accuracy can be further improved. Therefore, the trained MLP with 4000 samples are used in this work.

\subsection{BTE Thermal conductivity}

 \begin{figure}[h]
\includegraphics[width=0.5\linewidth]{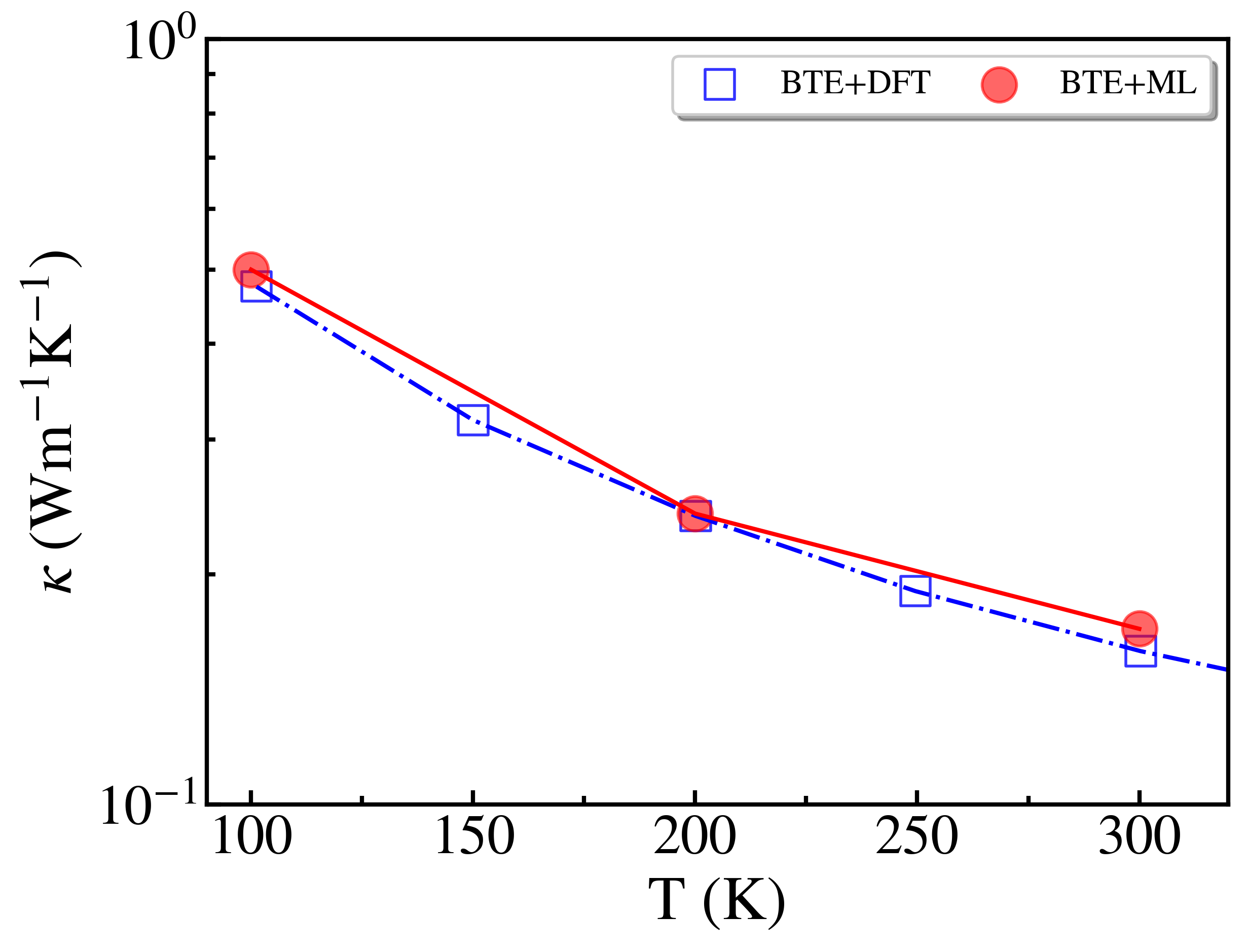}
\caption{The comparison of thermal conductivity predicted by DFT and MLP calculations versus temperature.
}
\label{figs3}
\end{figure}

We used VASP code \cite{Kresse1993,Kresse1996,KRESSE199615,Kresse1994} combined with the PHONONPY code \cite{phonopy} and the thirdorder.py \cite{Li2012} script to calculate the 2-rd and 3-rd interatomic force constants (IFCs) for Tl$_{3}$VSe$_{4}$. A set of supercells containing 2×2×2 conventional cells is simulated for the calculations of IFCs and the Monkhorst–Pack k-mesh of 5×5×5 is implemented to sample the irreducible Brillouin zone for the calculations of 2-rd and 3-rd IFCs, respectively. The cutoff energy is set to 550 eV. The IFCs are also calculated from the MLP with the same supercell and cutoff than the ones used in the DFT calculations of IFCs.

The phonon Boltzmann transport equation (PBTE) is then solved to calculate the thermal conductivity of Tl$_{3}$VSe$_{4}$ in the single-mode relaxation time approximation, as implemented in the ShengBTE software \cite{Li2014}. A q-grid of 16$\times$16$\times$16 is applied in the PBTE calculations. The comparison of thermal conductivities of DFT and MLP calculations, in which the IFCs are respectively calculated based on DFT and MLP interactions, is displayed in Fig.\,\ref{figs3}. The thermal conductivities from MLP calculations are well consistent with the results from DFT calculations in the full temperature region, indicating the high reliability of MLP in describing the phonon anharmonicity of Tl$_{3}$VSe$_{4}$. In addition, our MLP results also agree well with the predictions from Luo $et$ $al.$ \cite{Luo2020} that are based on the DFT interactions and without considering the temperature dependent effective potential.

\subsection{MD Thermal conductivity}

 \begin{figure}[h]
\includegraphics[width=0.8\linewidth]{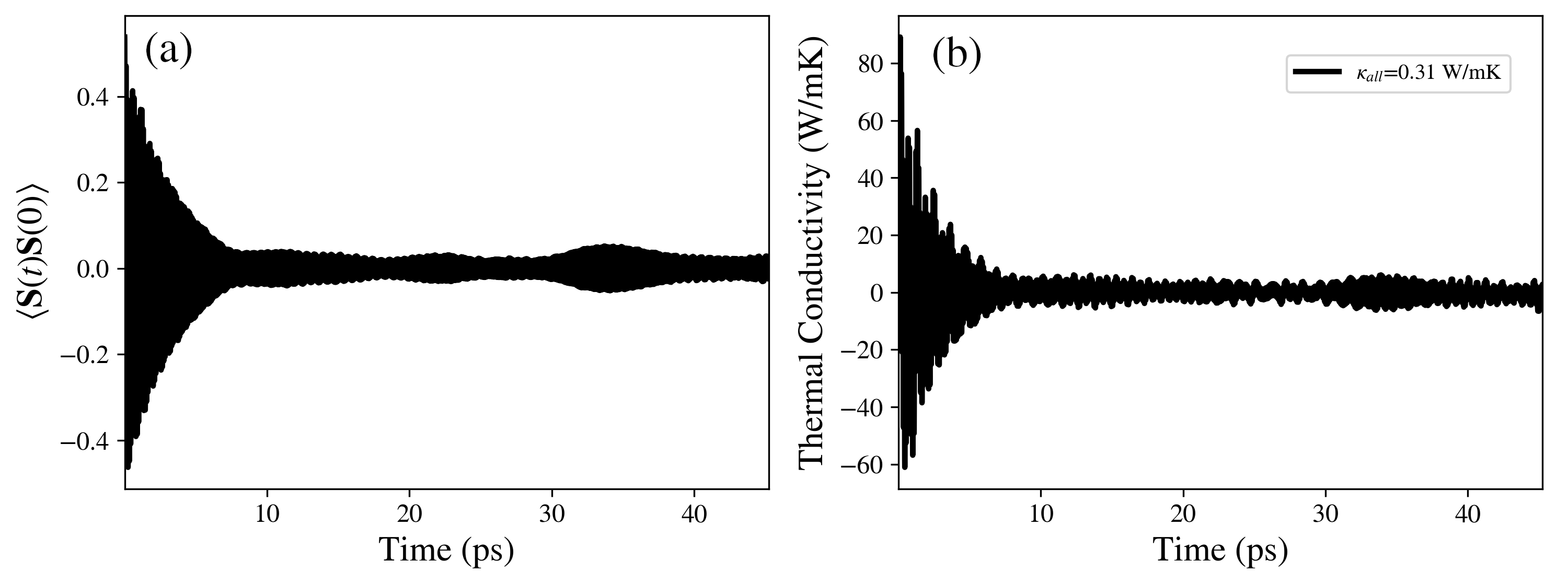}
\caption{The calculation of thermal conductivity based on the EMD simulations (a) The autocorrelation function of heat flux versus time. (b) The integral thermal conductivity from the Green-kubo approach versus integration time. The EMD simulations are carried out at room-temperature.
}
\label{figs4}
\end{figure}

All MD simulations are performed by using LAMMPS package \cite{Plimpton1995} with an interface to the MLP. The simulation size is chosen as $8\times 8\times 8$ the conventional cell. Periodic boundary conditions are applied in all directions. The time step is set as 0.15 fs in our simulations. After the structure relaxation and thermal equilibration in the isothermal-isobaric (NPT) ensemble for 30 ps, EMD simulations with the microcanonical (NVE) ensemble are performed for another 0.9 ns. The thermal conductivity of Tl$_{3}$VSe$_{4}$ is also calculated during the EMD simulations based on the Green-Kubo approach in Eq. (1) of the maintext. Here, the heat current is calculated as \cite{Torii2008}:

\begin{eqnarray}
\mathbf{S}=\frac{1}{V}\left [ \sum_{i}^{N}E_{i}\mathbf{v} _{i} +\frac{1}{2}\sum_{i}^{N}\sum_{j> i}^{N}\left ( \mathbf{F}_{ij}\cdot \left ( \mathbf{v} _{i}+\mathbf{v} _{j} \right ) \right )\mathbf{r}_{ij}\right ],
\label{eqs1}
\end{eqnarray}

\noindent where $E_{i}$ and $\mathbf{v} _{i}$ are the total energy and velocity, respectively, for the $i$-th atom. $\mathbf{r}_{ij}$ and $\mathbf{F}_{ij}$ are the distance and force between two atoms $i$ and $j$, respectively. For each case, 20 independent runs are performed in order to obtain a stable averaged of $\kappa$. The correlation time considered in our simulation is long enough to ensure the proper decaying of heat current autocorrelation function as displayed in Fig.\,\ref{figs4}. The isotope scattering effect is included by randomly change the mass of atoms in MD simulations.

\section{Wavelet transform}
\subsection{Approach}

The temporal coherence of thermal phonons can be defined in the following basis

\begin{eqnarray}
\centering
\psi_{\omega_{\lambda} ,t_{0},\Delta_{\lambda} } \left ( t \right )=\pi ^{-\frac{1}{4}}\Delta_{\lambda} ^{-\frac{1}{2}}e^{\left [ i\omega_{\lambda} \left ( t-t_{0} \right ) \right ]}e^{\left [ -\frac{1}{2}\left ( \frac{t-t_{0}}{\Delta_{\lambda}} \right) ^{2} \right ]},
\label{eqs2}
\end{eqnarray}

\noindent where $\omega_{\lambda}$ is the angular frequency of mode ${\lambda} $, and $\Delta_{\lambda} $ defines the wavepacket duration. $t$ corresponds to the time variable, and $t_{0}$ to the position of highest amplitude in the wavepacket and also corresponds to the time evolution in the wavelet space. Inside the wavepacket, planewaves are in phase, the $\Delta_{\lambda}$ term in Eq. (\ref{eqs2}) is thus a measure of the temporal coherence of thermal phonons. Here, we define the wavepacket full-width at half-maximum (FWHM) as the coherence time $\tau_{\lambda}^{c}=2\sqrt{2ln2}\Delta_{\lambda}$. The temporal coherence information of thermal phonons can be calculated from the wavelet transform as,

\begin{eqnarray}
\Lambda \left ( \omega_{\lambda},t_{0},\tau_{\lambda}^{c}\right )=\int \psi_{\omega_{\lambda},t_{0},\tau_{\lambda}^{c} } \left ( t \right )F\left (  t_{0}\right )dt,
\label{eqs3}
\end{eqnarray}

\noindent where $F\left (  t_{0}\right )$ denotes the time dependent dynamical quantity, which is chosen as the phonon modal velocity $\mathbf{\upsilon} \left ( \mathbf{k},s \right )$ \cite{Larkin2014a}.

\begin{eqnarray}
\mathbf{\upsilon} \left ( \mathbf{k},s \right ) =\frac{1}{a}\sum_{b,l}^{a}\left [ \mathbf{\dot{u}}_{bl}\left ( t \right )\cdot \mathbf{e}^{\ast } _{b}\left ( \mathbf{k},s \right )\times exp\left ( i\mathbf{k}\cdot \mathbf{R}_{0l} \right )\right ],
\label{eqs4}
\end{eqnarray}

\noindent where $\mathbf{\dot{u}}_{bl}\left ( t \right )$ is the velocity of the $b$th atom in the $l$th unit cell at time $t$, $a$ is the number of cell, $\mathbf{e}^{\ast } \left ( \lambda \right )$ the complex conjugate of the eigenvector of mode $\lambda$, and $\mathbf{R}_{0l} $ is the equilibrium position of the $l$th unit cell. Here, $\mathbf{k}$ and $s$ corresponding to the mode $\lambda$. At the frequency $\omega_{\lambda}$, the time dependent phonon number of a given coherence time, here called phonon number density, $N\left ( t_{0},\tau _{\lambda}^{c}  \right )$ can be calculated as $N\left ( t_{0},\tau _{\lambda}^{c}  \right )=\frac{1}{2}m\left | \Lambda \left ( \omega_{\lambda},t_{0},\tau _{\lambda}^{c}  \right ) \right |^{2}/\hbar\omega_{\lambda}$. 
 
\subsection{Coherence}

 \begin{figure}[h]
\includegraphics[width=0.5\linewidth]{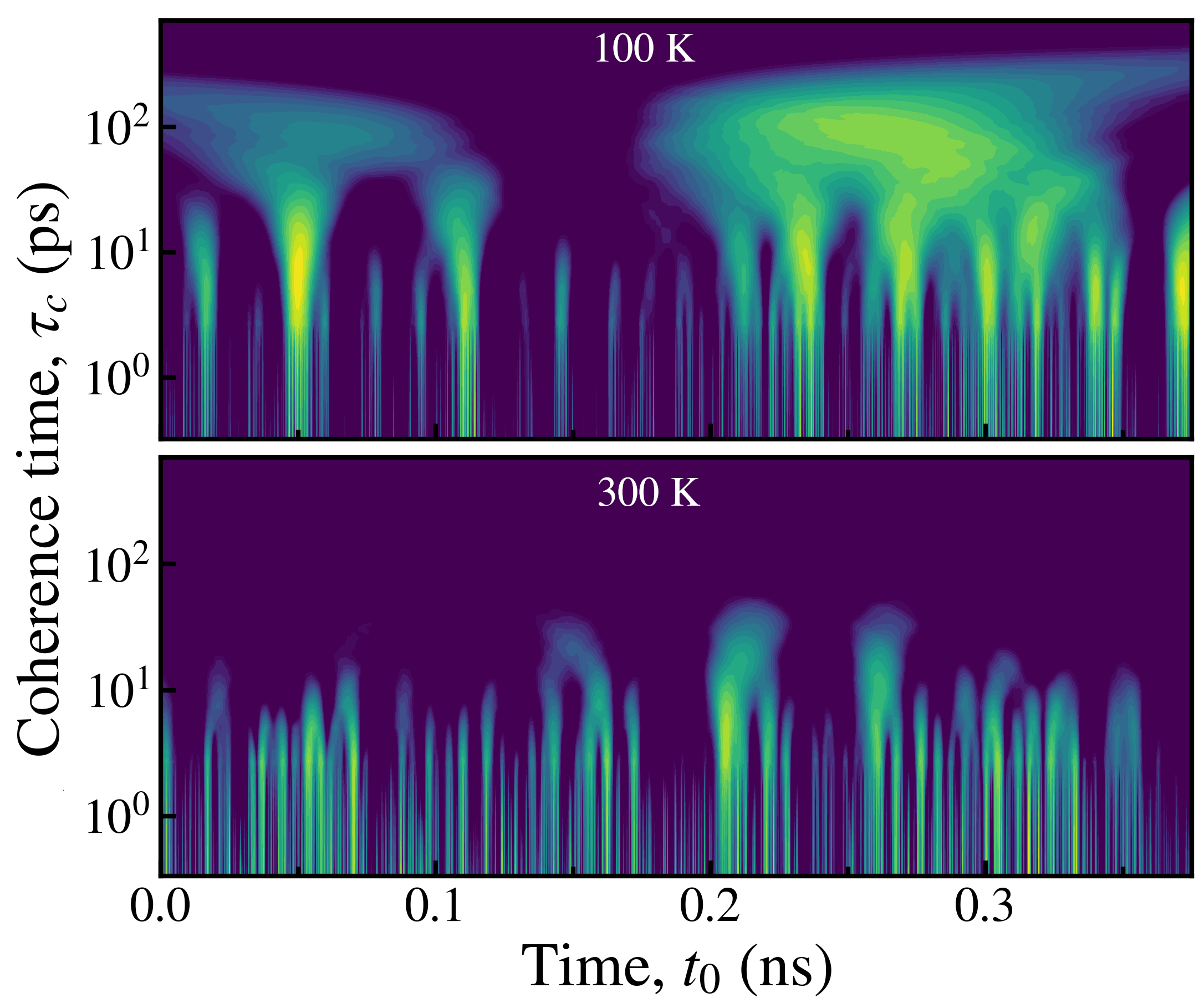}
\caption{Evolution time and coherence time dependent phonon number of Tl$_{3}$VSe$_{4}$ for the 0.25 THz mode at 100K and 300K.
}
\label{figs5}
\end{figure}

Fig.\,\ref{figs4} shows the phonon number density as a function of evolution time and coherence time for a low frequency mode, i.e., 0.25 THz. As increasing of temperature from 100K to 300K, the coherence time of this mode is significantly suppressed, indicating the dephasing of phonons due to the enhanced phonon-phonon scattering.

\section{Phonon decay}

\subsection{Correlation function}

The wavepackets are distributed along the coherence coherence time and can be further investigated by building the time-averaged phonon number density versus coherence time 

\begin{eqnarray}
D\left (\omega _{\lambda},\tau _{\lambda}^{c} \right )=\frac{1}{N_{t_{0}}}\sum_{t_{0}}\frac{N\left ( \omega_{\lambda},t_{0},\tau _{\lambda}^{c} \right )}{\sum_{\tau _{\lambda}^{c}}N\left ( \omega_{\lambda},t_{0},\tau _{\lambda}^{c} \right )}
\label{eqs5}
\end{eqnarray}

\noindent where $N_{t_{0}}$ denotes the number of terms in the sum. Then, the phonon decay can be calculated from the correlation function

\begin{eqnarray}
Cor\left ( t,\tau _{\lambda}^c \right )=\frac{\left \langle {\Delta N}\left (t,\tau _{\lambda}^c   \right ) {\Delta N}\left (0,\tau _{\lambda}^c \right ) \right \rangle}{\left \langle {\Delta N}\left (0,\tau _{\lambda}^c \right ) {\Delta N}\left (0,\tau _{\lambda}^c \right ) \right \rangle}. 
\label{eqs6}
\end{eqnarray}

\noindent where, ${\Delta N}(t_{0},\tau _{\lambda}^c)=N\left ( t_{0},\tau _{\lambda}^c \right )-\left \langle N(t_{0},\tau _{\lambda}^c) \right \rangle_{t_{0}}$. To obtain a coherence time independent phonon decay function, an average over coherence times can be further implemented

\begin{eqnarray}
Cor\left ( t \right )_{\lambda}=\sum_{\tau _{\lambda}^{c}}D\left (\tau _{\lambda}^{c} \right )Cor\left ( t,\tau _{\lambda}^c \right ).
\label{eqs7}
\end{eqnarray}

\subsection{Coherence time and Lifetime}

The phonon decays are as numerous as the number of coherence times. As input to the general thermal conductivity expression (Eq. (6) in maintext), we define the averaged coherence time ($\bar{\tau}_{\lambda}^{c}$) 

\begin{eqnarray}
\bar{\tau}_{\lambda}^{c}=\sum_{\tau _{\lambda}^{c}}D\left (\tau _{\lambda}^{c} \right )\tau _{\lambda}^{c}.
\label{eqs8}
\end{eqnarray}

Mean lifetimes ($\bar{\tau}_{\lambda}^{p}$) are obtained by fitting the averaged phonon decay $Cor\left ( t \right )_{\lambda}$ as
  
\begin{eqnarray}
Cor\left ( t \right )_{\lambda}= e^{-\frac{t}{2\bar{\tau}_{\lambda}^{p} }}e^{-4ln2\frac{t^{2}}{{\bar{\tau}_{\lambda}}^{c2}}}.
\label{eqs9}
\end{eqnarray}

\begin{figure}[h]
\includegraphics[width=0.55\linewidth]{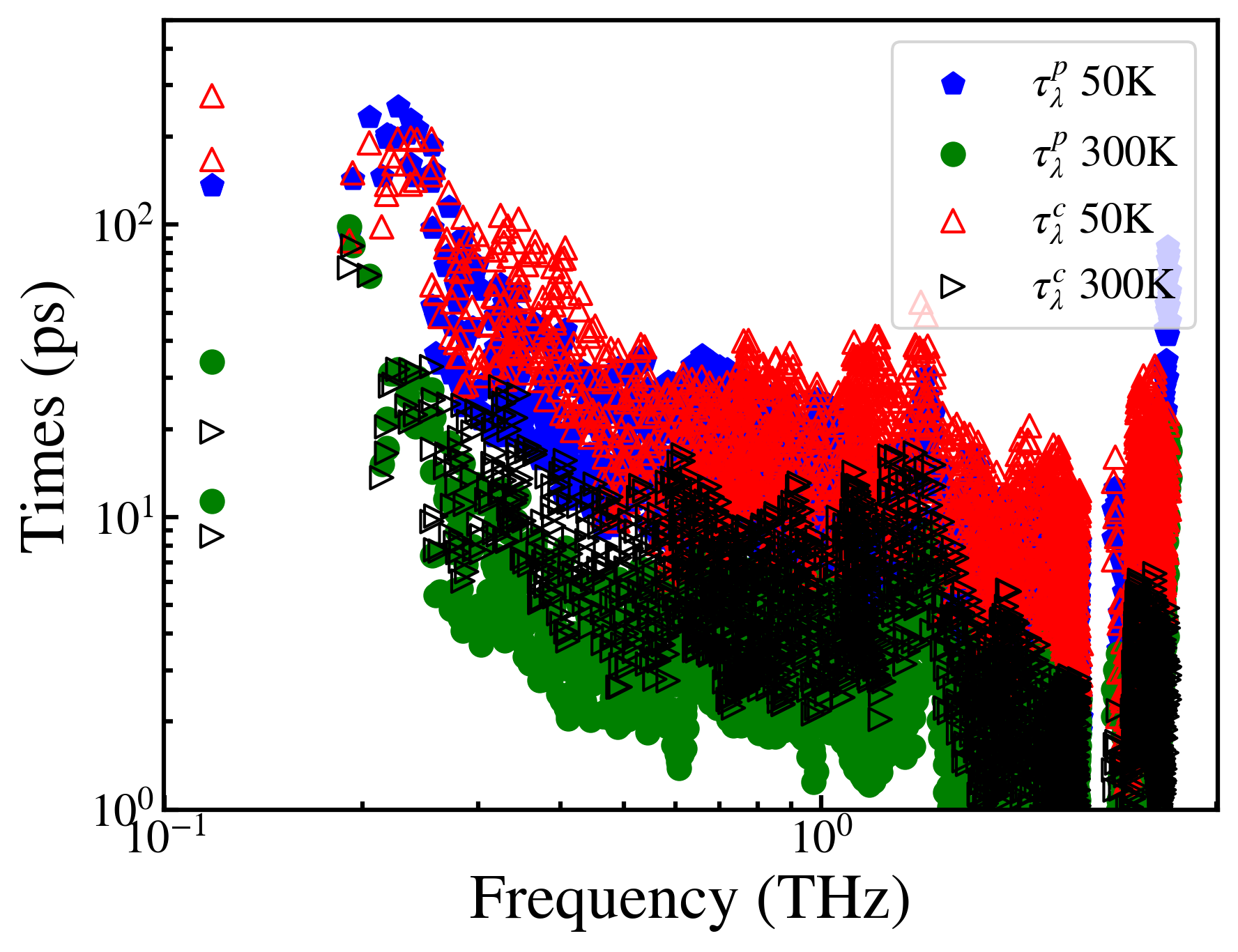}
\caption{The calculated coherence times and lifetimes versus frequency at 50K and 300K.
}
\label{figs6}
\end{figure}

In the maintext, especially in the thermal conductivity model, $\tau_{\lambda}^{p}$ and $\tau_{\lambda}^{c}$ refer respectively to $\bar{\tau}_{\lambda}^{p}$ and $\bar{\tau}_{\lambda}^{c}$ of above formulas. The fitted $\tau_{\lambda}^{p}$ and $\tau_{\lambda}^{c}$ are reported in Fig.\,\ref{figs6}.

\section{Deviation}

 \begin{figure}[h]
\includegraphics[width=0.55\linewidth]{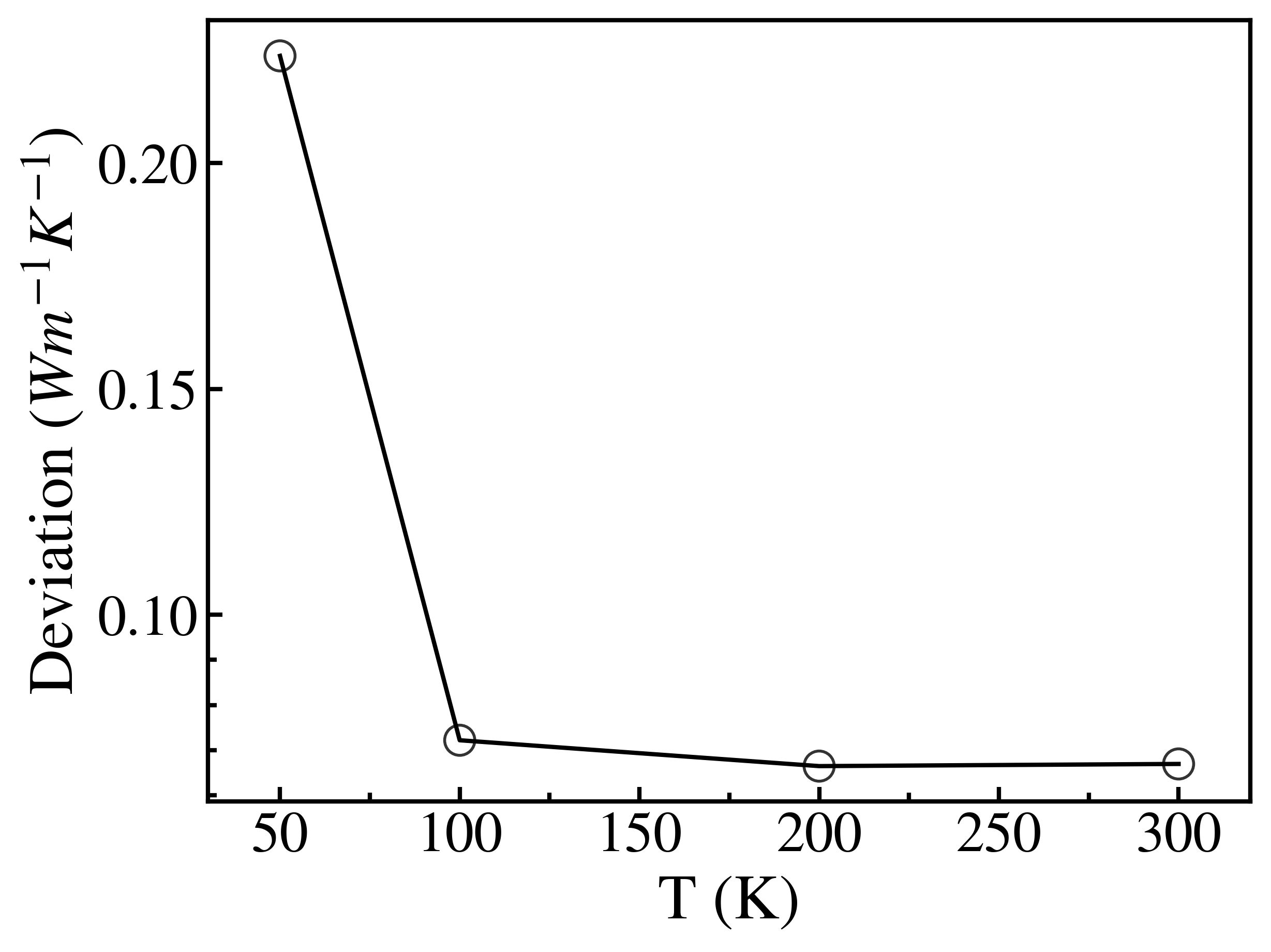}
\caption{The deviation of the reported thermal conductivities.
}
\label{figs7}
\end{figure}

The deviation (i.e. difference) of thermal conductivities between the reported values in \cite{Mukhopadhyay2018} and \cite{Jain2020}, suggesting that the coherent theory from \cite{RN1616} fails to capture the intrinsic coherence at low temperatures is reported in Fig.\,\ref{figs7}.

~\\
~\\
~\\

\bibliographystyle{apsrev4-2}
\bibliography{library}